\documentclass[%
reprint,
superscriptaddress,
 amsmath,amssymb,
pra,
]{revtex4-2}

\usepackage{amsmath}
\usepackage {xcolor}
\usepackage{comment}
\usepackage{mathrsfs}
\usepackage{comment}
\usepackage{subfigure}
\usepackage[colorlinks,
            linkcolor=blue,
            anchorcolor=blue,
            citecolor=blue]{hyperref}
\newtheorem{theorem}{Theorem}

\def\ra{\rangle}
\def\la{\langle}
\allowdisplaybreaks[4]
\usepackage{graphicx}
\usepackage{dcolumn}
\usepackage{bm}

\begin{document}


\title{Parameterized steering criteria via correlation matrices}

\author{Qing-Hua Zhang}
\email[]{qhzhang@csust.edu.cn}

\affiliation{School of Mathematics and Statistics, Changsha University of Science and Technology, Changsha 410114, China}

\author{Lemin Lai}
\email[]{lai$_$lm@amss.ac.cn}

\affiliation{Academy of Mathematics and Systems Science, Chinese Academy of Sciences, Beijing 100190, China}
\affiliation{School of Mathematical Sciences, University of Chinese Academy of Sciences, Beijing 100049, China}

\author{Shao-Ming Fei}
\email[]{feishm@cnu.edu.cn}
\affiliation{School of Mathematical Sciences, Capital Normal University,
Beijing 100048, China}
\affiliation{Max-Planck-Institute for Mathematics in the Sciences, 04103 Leipzig, Germany}

\begin{abstract}
We study the steerability for arbitrary dimensional bipartite systems based on the correlation matrices given by local special unitary groups. We present families of steering criteria for bipartite quantum states in terms of parameterized correlation matrices. We show that these steering criteria may detect more steerable states than the existing steering criteria. The results are illustrated by detailed examples.
\end{abstract}

\maketitle

\section{Introduction}

The well-known Einstein-Podolsky-Rosen (EPR) paradox reveals counterintuitive features of quantum mechanics~\cite{einstein1935can,schrodinger1935discussion}. Quantum steering, as a correlation intermediate between Bell nonlocality~\cite{brunner2014bell} and quantum entanglement~\cite{horodecki2009quantum}, characterizes the ability of Alice to remotely affect or steer Bob's local state by Alice's arbitrary choice of measurement setting~\cite{simon2001no}. Wiseman $et\ al.$ rigorously formulated an operational definition of quantum steering in terms of local hidden state (LHS) models~\cite{wiseman2007steering}. In contrast of Bell nonlocality and entanglement, quantum steering demonstrates asymmetric behaviour in which one party can steer another party, but not always the vice versa~\cite{midgley2010asymmetric,reid2013monogamy, bowles2014one,bowles2016sufficient}. Moreover, not every entangled state exhibits the steerability, and not every steerable state violates a Bell inequality~\cite{wiseman2007steering}. Besides its foundational significance, EPR-steering has a wide range of applications in many quantum information processing tasks such as one-sided device-independent quantum key distribution~\cite{branciard2012one,gehring2015implementation,walk2016experimental}, quantum secret sharing~\cite{xiang2017multipartite}, quantum networking tasks~\cite{huang2019securing,cavalcanti2015detection,armstrong2015multipartite}, subchannel discrimination~\cite{piani2015necessary,sun2018demonstration}, quantum teleportation~\cite{he2015secure,reid2013signifying} and randomness certification~\cite{passaro2015optimal,skrzypczyk2018maximal,mattar2017experimental}.

A natural problem about EPR steering is to identify whether a given quantum state is steerable. Different EPR steering criteria have been derived, including linearity~\cite{PRXQuantum.3.030102,cavalcanti2009experimental,zheng2017efficient,saunders2010experimental,RevModPhys.92.015001,PhysRevLett.128.240402,zheng2016certifying}, uncertainty relations~\cite{PhysRevLett.106.130402,PhysRevA.87.062103,PhysRevA.92.062130,PhysRevA.92.062130,PhysRevA.98.050104,PhysRevA.98.062111,PhysRevA.101.022324,PhysRevA.105.022421}, covariance matrices~\cite{lai2022detecting}, Clauser-Horne-Shimony-Holt-type inequalities~\cite{Cavalcanti:15,PhysRevA.94.032317,PhysRevA.95.062111,Fan:23}, detection of entanglement \cite{PhysRevA.101.042115,PhysRevA.99.052109,PhysRevA.98.052114,https://doi.org/10.1002/andp.202100098},moment matrices~\cite{PhysRevLett.115.210401} and machine learning~\cite{PhysRevA.100.022314,PhysRevA.104.052427,PhysRevA.105.032408}. However, more comprehensive and operational steering criteria particularly for higher-dimensional systems are still far from being satisfied, which may strengthen the understanding of quantum correlations and extend potential applications.

Let Alice and Bob be two spatially separated observers. The local sets of observables (measurements) of Alice and Bob, are denoted by $\mathbf{A}=\{A_i\}$ and $\mathbf{B}=\{B_j\}$, respectively, with $a\, (b)$ labeling the measurement outcomes of $A_i \,(B_j)$. Alice prepares a bipartite state $\rho$ and sends a party to Bob. Alice's task is to convince Bob that they share an entangled state. In the steering task, Bob does not trust Alice and will be convinced only if he verifies that Alice can steer his state. Bob asks Alice to carry out a measurement $A_i=\{\Pi_a^{A_i}\}$ with measurement operators $\Pi_a^{A_i}$, and to announce her measurement outcomes by classical communication. Then Bob performs a measurement $B_j=\{\Pi_b^{B_j}\}$ on his party. The joint probability distribution is given by
\begin{equation}
P(a,b|A_i,B_j;\rho)=\mathbf{tr}(\Pi_a^{A_i}\otimes\Pi_b^{B_j}\rho).
\end{equation}
Alice will fail to convince Bob that she can steer Bob's state if there exists an ensemble $\{(p_\xi,\rho_\xi)\}$ on Bob's reduced state such that
\begin{equation}
P(a,b|A_i,B_j;\rho)=\sum_\xi p_\xi P(a|A_i;\xi)P_Q(b|B_j;\rho_\xi).
\end{equation}
for all $A_i\in\mathbf{A}$ and $B_j\in\mathbf{B}$, where $P(a|A_i;\xi)$ represents joint probability distribution of measurement $A_i$ and the predetermined local hidden variable (LHV) $\xi$, $P_Q(b|B_j;\rho_\xi)=\mathbf{tr}(\Pi_b^{B_j}\rho_\xi)$ is the probability of outcome $b$ given by measuring $B_j$ on the predetermined local hidden state (LHS) $\rho_\xi$. In this scenario, $\sigma_{a|A_i}=\mathbf{tr}_A(\Pi_{a}^{A_i}\otimes I\rho)$ is the (un-normalized) conditional state held by Bob after Alice performed the measurement $A_i$ with outcome $a$. If Alice can not steer Bob's state, the following LHS model holds,
\begin{equation}
\sigma_{a|A_i}=\sum_\xi p_\xi P(a|A_i)\rho_\xi
\end{equation}
for all measurement $A_i\in \mathbf{A}$~\cite{wiseman2007steering}.

In this work, we derive stronger steering criteria based on parameterized correlation matrices of local observables which apply to arbitrary dimensional bipartite quantum systems. This approach is an extension from the entanglement detection via correlation matrices~\cite{shen2016improved,zhu2023family,PhysRevA.101.012341}. Several examples are employed to illustrate the performance of our steering criteria.

\section{Parameterized steering criteria}

In this section, we present a family of parameterized steering criteria based on correlation matrices. Consider a bipartite state $\rho$ shared by Alice and Bob on a Hilbert space $H_A\otimes H_B$ with dimension $d_A\times d_B$. Let $\{G_i\}_{i=1}^{{d^2_A-1}}$ and $\{H_j\}_{j=1}^{{d^2_B-1}}$ be local generators of the special unitary groups $SU(d_A)$ and $SU(d_B)$, respectively, with relations $\mathbf{tr}(G_iG_j)=\delta_{ij}$, $\mathbf{tr}(G_i)=0$, $\sum_i G_i^2=d_A\mathbb{I}_A-1/d_A\mathbb{I}_A$ with $\mathbb{I}_A$ the identity operator on $H_A$, $\sum_i \mathbf{tr}(G_i\rho_A)^2=\mathbf{tr}(\rho_A^2)-1/d_A$ and similarly for $\{H_j\}_{j=1}^{{d^2_B-1}}$. Any bipartite quantum state $\rho$ can be written as
$$
\begin{aligned}
\rho= & \frac{\mathbb{I}_A}{d_A} \otimes \frac{\mathbb{I}_B}{d_B}+\sum_{i>0} r_i G_i \otimes \frac{\mathbb{I}_B}{d_B}+\sum_{j>0} s_j \frac{\mathbb{I}_A}{d_A} \otimes H_j \\
& +\sum_{i, j>0} t_{i j} G_i \otimes H_j,
\end{aligned}
$$
where $r_i=\mathbf{tr} (\rho G_i \otimes \mathbb{I}_B)$ and $s_j=\mathbf{tr} (\rho {\mathbb{I}_A}\otimes H_j)$ are the elements of generalized Bloch vectors corresponding to the reduces states $\rho_A$ and $\rho_B$, respectively, and $T_\rho=(t_{i j})$ is usual correlation tensor with $t_{i j}=\mathbf{tr} (\rho G_i \otimes H_j)$. The reduced states are of forms,
$$
\rho_A=\mathbf{tr}_B \rho=\frac{\mathbb{I}_A}{d_A}+\sum_{i>0} r_i G_i,
$$
and
$$
\rho_B=\mathbf{tr}_A \rho=\frac{\mathbb{I}_B}{d_B}+\sum_{j>0} s_j H_j.
$$

Set
$$r=(r_1, r_2,\cdots, r_{d_A^2-1})^T,$$ and $$s=(s_1, s_2,\cdots, s_{d_B^2-1})^T.$$
Motivated by the approach used in entanglement criteria \cite{shen2016improved,zhu2023family,PhysRevA.101.012341}, we define a parameterized correlation matrix,
\begin{equation}
\tilde{T}_\rho^{\alpha,\beta}=
\begin{pmatrix}
      \alpha\beta^T& \alpha s^{T}    \\
     r\beta^T&  T_\rho
\end{pmatrix},
\end{equation}
where $\alpha=(\alpha_1,\alpha_2,\cdots,\alpha_n)^T$ and $\beta=(\beta_1,\beta_2,\cdots,\beta_m)^T$ are arbitrary $n$ and $m$-dimensional vectors, with $\alpha_i$ and $\beta_j$ being nonnegative real numbers.
Let $\|M\|_{\rm tr}=\mathbf{tr}(\sqrt{M^\dagger M})$ denote the trace norm of the matrix $M$, and $\|\alpha\|=\sqrt{\sum_{i=1}^n \alpha_i^2}$ be the spectral norm of $\alpha$.

\begin{theorem}\label{th1}
If the state $\rho$ is not steerable from Alice to Bob, we have
\begin{equation}
\|\tilde{T}_\rho^{\alpha,\beta}\|_{\rm tr}\leqslant \sqrt{\|\alpha\|^2+d_A-\frac{1}{d_A}}\sqrt{\|\beta\|^2+1-\frac{1}{d_B}}
\end{equation}
for all $\alpha$ and $\beta$.
\end{theorem}

\textit{Proof.} If the state $\rho$ is not steerable from Alice to Bob, there exists an ensemble $\{p_\xi,\rho_\xi\}$ of Bob's reduced states such that
\begin{align*}
\mathbf{tr}((A \otimes B) \rho) & =\sum_{a, b} ab P(a, b | A, B ; \rho) \\
& =\sum_{a, b} ab\sum_{\xi} p_{\xi} P(a| A ; \xi) P_Q(b | B ; \rho_{\xi})\\
& =\sum_{\xi} p_{\xi} \sum_a a P(\alpha | A ; \xi) \sum_b b P_Q (b | B ; \rho_{\xi}) \\
& =\sum_{\xi} p_{\xi}\langle A\rangle_{\xi} \mathbf{tr} (B \rho_{\xi}),
\end{align*}
where $\langle A\rangle_{\xi} =\sum_a a P(a | A;\xi)$. With respect to the generators $\{G_i\}_{i=1}^{{d^2_A-1}}$ and $\{H_j\}_{j=1}^{{d^2_B-1}}$, the Bloch vectors and the correlation tensors can be written as
\begin{align*}
r_i & =\mathbf{tr}(G_i \rho_a)=\sum_{\xi} p_{\xi}\langle G_i\rangle_{\xi},\\
s_j & =\mathbf{tr}(H_j \rho_b)=\sum_{\xi} p_{\xi} \mathbf{tr}(H_j \rho_{\xi}),\\
t_{i j} & =\sum_{\xi} p_{\xi}\langle G_i\rangle_{\xi} \mathbf{tr}(H_j\rho_{\xi}).
\end{align*}

Define vectors $\mu=(\langle G_1\rangle_{\xi}, \langle G_2\rangle_{\xi},\cdots,\langle G_{d_A^2-1}\rangle_{\xi})^T$ and $\nu=(\mathbf{tr}(H_1 \rho_{\xi}), \mathbf{tr}(H_2 \rho_{\xi}),\cdots, \mathbf{tr}(H_{d_B^2-1}\rho_{\xi}))^T$. We have
\begin{equation}\label{munu}
\tilde{T}_\rho^{\alpha,\beta}=\sum_\xi p_\xi
\begin{pmatrix}
      \alpha\beta^T& \alpha \nu^{T}    \\
     \mu\beta^T&  \mu\nu^T
\end{pmatrix}
=\sum_\xi p_\xi
\begin{pmatrix}
      \alpha   \\
     \mu
\end{pmatrix}
\begin{pmatrix}
      \beta^T&\nu^T
\end{pmatrix},
\end{equation}
from which we get
\begin{align*}
\|\tilde{T}_\rho^{\alpha,\beta}\|_{\rm tr}\leqslant& \sum_\xi p_\xi
\left\|
\begin{pmatrix}
      \alpha   \\
     \mu
\end{pmatrix}
\right\|_{\rm tr}
\left\|
\begin{pmatrix}
      \beta^T&\nu^T
\end{pmatrix}
\right\|_{\rm tr}       \\
\leqslant& \sqrt{\sum_\xi p_\xi
\left\|
\begin{pmatrix}
      \alpha   \\
     \mu
\end{pmatrix}
\right\|^2}\sqrt{\sum_\xi p_\xi
\left\|
\begin{pmatrix}
      \beta^T&\nu^T
\end{pmatrix}
\right\|^2}          \\
=&\sqrt{\|\alpha\|^2+\sum_\xi p_\xi\|\mu\|^2}\sqrt{\|\beta\|^2+\sum_\xi p_\xi\|\nu\|^2}\\
=&\sqrt{\|\alpha\|^2+\sum_\xi p_\xi\sum_i\la G_i\ra_\xi^2}\\
&\times\sqrt{\|\beta\|^2+\sum_\xi p_\xi\sum_j \mathbf{tr}(H_j\rho_\xi)^2}\\
\leqslant&\sqrt{\|\alpha\|^2+\sum_\xi p_\xi\sum_i\la G_i^2\ra_\xi}\\
&\times\sqrt{\|\beta\|^2+\sum_\xi p_\xi\sum_j \mathbf{tr}(H_j\rho_\xi)^2}\\
\leqslant&\sqrt{\|\alpha\|^2+d_A-\frac{1}{d_A}}\sqrt{\|\beta\|^2+1-\frac{1}{d_B}},
\end{align*}
where the first inequality is due to the convexity of trace norm, the second inequality follows from the Cauchy Schwarz inequality, the third inequality comes from $\la A^2\ra_\xi\geqslant \la A\ra_\xi^2$ with $\langle A^2\rangle_{\xi} =\sum_a a^2 P(a | A;\xi)$.
This completes the proof. $\Box$

Note that the trace norm $\|\tilde{T}_\rho^{\alpha,\beta}\|_{\rm tr}=\mathbf{tr} (\sqrt{(\tilde{T}_\rho^{\alpha,\beta})^\dagger\tilde{T}_\rho^{\alpha,\beta}})$, where
\begin{equation}\label{xy}
\begin{aligned}
(\tilde{T}_\rho^{\alpha,\beta})^\dagger\tilde{T}_\rho^{\alpha,\beta}=\left(
\begin{array}{cc}
(\|\alpha\|^2+\|r\|^2) \beta \beta^T & \beta (\|\alpha\|^2 s^T+r^T T_\rho) \\
(\|\alpha\|^2 s+T_\rho^\dagger r) \beta^T & \|\alpha\|^2 s s^T+T_\rho^\dagger T_\rho
\end{array}\right).
\end{aligned}
\end{equation}
According to Lemma 1 in Ref.~\cite{zhu2023family}, the singular values of $\tilde{T}_\rho^{\alpha,\beta}$ are same for any different parameter pairs $(\alpha_1,\beta_1)$ and $(\alpha_2,\beta_2)$ such that $\|\alpha_1\|=\|\alpha_2\|$ and $\|\beta_1\|=\|\beta_2\|$. Thus,  Theorem \ref{th1} is equivalent to
\begin{equation}
\|\tilde{T}_\rho^{x,y}\|_{\rm tr}\leqslant \sqrt{x^2+d_A-\frac{1}{d_A}}\sqrt{y^2+1-\frac{1}{d_B}},
\end{equation}
where $x$ and $y$ are nonnegative real numbers. The Theorem \ref{th1} corresponds to the steering criteria in Ref.~\cite{lai2023steerability} which is based on the Heisenberg-Weyl observable basis.

Furthermore, consider any two positive real numbers $x$ and $y$, two real vectors $\eta=(\eta_1,\eta_2,\cdots,\eta_{n^\prime})^T$ and $\gamma=(\gamma_1,\gamma_2,\cdots,\gamma_{m^\prime})^T$ with nonnegative real elements. We define the following new parameterized correlation matrix,
\begin{equation}
\tilde{T}_\rho^{x,y,\eta,\gamma}=
\begin{pmatrix}
      xy& x\gamma^T\otimes s^T    \\
    y \eta\otimes r&  \eta\gamma^T\otimes T_\rho
\end{pmatrix}.
\end{equation}

\begin{theorem}\label{th2} If the state $\rho$ is not steerable from Alice to Bob, we have
\begin{equation}
\|\tilde{T}_\rho^{x,y,\eta,\gamma}\|_{\rm tr}\leqslant \sqrt{x^2+\|\eta\|^2(d_A-\frac{1}{d_A})}\sqrt{y^2+\|\gamma\|^2(1-\frac{1}{d_B})}.
\end{equation}
\end{theorem}

\textit{Proof.} Similar to the proof of Theorem \ref{th1}, if Bob admits the LHS models, there exists an ensemble $\{p_\xi,\rho_\xi\}$ of Bob's reduced states such that
\begin{align*}
\|\tilde{T}_\rho^{x,y,\eta,\gamma}\|_{\rm tr}=& \left\|\sum_\xi p_\xi
\begin{pmatrix}
      x   \\
     \eta\otimes\mu
\end{pmatrix}
\begin{pmatrix}
      y&\gamma^T\otimes\nu^T
\end{pmatrix}
\right\|_{\rm tr}       \\
\leqslant& \sum_\xi p_\xi
\left\|
\begin{pmatrix}
      x   \\
     \eta\otimes\mu
\end{pmatrix}
\right\|_{\rm tr}
\left\|
\begin{pmatrix}
      y&\gamma^T\otimes\nu^T
\end{pmatrix}
\right\|_{\rm tr}       \\
\leqslant& \sqrt{\sum_\xi p_\xi
\left\|
\begin{pmatrix}
      x   \\
     \eta\otimes\mu
\end{pmatrix}
\right\|^2}\sqrt{\sum_\xi p_\xi
\left\|
\begin{pmatrix}
      y&\gamma^T\otimes\nu^T
\end{pmatrix}
\right\|^2}          \\
=&\sqrt{x^2+\|\eta\|^2\sum_\xi p_\xi\|\mu\|^2}\sqrt{y
^2+\|\gamma\|^2\sum_\xi p_\xi\|\nu\|^2}\\
\leqslant&\sqrt{x^2+\|\eta\|^2(d_A-\frac{1}{d_A})}\sqrt{y
^2+\|\gamma\|^2(1-\frac{1}{d_B})},
\end{align*}
where $\mu$ and $\nu$ are the same as in (\ref{munu}). $\Box$

Note that
\begin{equation}\label{xygh}
\begin{aligned}
&(\tilde{T}_\rho^{x,y,\eta,\gamma})^\dagger \tilde{T}_\rho^{x,y,\eta,\gamma}\\=&\left(
\begin{array}{cc}
x^2 y^2+y\|\eta\|^2\|r\|^2 & y \gamma^T \otimes\left(x^2 s^T+\|\eta\|^2 r^T T_\rho\right) \\ y \gamma \otimes\left(x^2 s+\|\eta\|^2 T_\rho^\dagger r\right) & \gamma \gamma^T \otimes\left(x^2 s s^T+\|\eta\|^2 T_\rho^\dagger T_\rho\right)
\end{array}\right).
\end{aligned}
\end{equation}
When one takes $\|\eta_1\|=\|\eta_2\|$, then the trace norms of $\|\tilde{T}_\rho^{x,y,\eta,\gamma}\|_{\rm tr}$ are invariant, $i.e.,$ $\|\tilde{T}_\rho^{x,y,\eta_1,\gamma}\|_{\rm tr}=\|\tilde{T}_\rho^{x,y,\eta_2,\gamma}\|_{\rm tr}$. Since $\|G\|_{\rm tr}=\mathbf{tr}(\sqrt{G^\dagger G})=\mathbf{tr}(\sqrt{GG^\dagger})$, symmetrically one has $\|\tilde{T}_\rho^{x,y,\eta,\gamma_1}\|_{\rm tr}=\|\tilde{T}_\rho^{x,y,\eta,\gamma_2}\|_{\rm tr}$. Let $g$ and $h$ be two nonnegative real numbers, our Theorem~\ref{th2} is then equivalent to
\begin{equation}\label{steerxygh}
\|\tilde{T}_\rho^{x,y,g,h}\|_{\rm tr}\leqslant \sqrt{x^2+g^2(d_A-\frac{1}{d_A})}\sqrt{y^2+h^2(1-\frac{1}{d_B})}.
\end{equation}

More generally, one may replace $x,y$ by two vectors $\alpha,\beta$. Then the correlation tensor matrix is of the form,
\begin{equation}
\tilde{T}_\rho^{\alpha,\beta,\eta,\gamma}=
\begin{pmatrix}
      \alpha\beta^T& \alpha\gamma^T\otimes s^T    \\
    \eta\otimes r\beta^T&  \eta\gamma^T\otimes T_\rho
\end{pmatrix}.
\end{equation}
Similar to the proofs of Theorems \ref{th1} and \ref{th2}, if Alice can not steer Bob's state, we have
\begin{equation}\label{th3}
\begin{aligned}
\tilde{T}_\rho^{\alpha,\beta,\eta,\gamma} \leqslant& \sqrt{\|\alpha\|^2+\|\eta\|^2(d_A-\frac{1}{d_A})}\\
&\times\sqrt{\|\beta\|^2+\|\gamma\|^2(1-\frac{1}{d_B})}.
\end{aligned}
\end{equation}
Remarkably, according to the discussion for (\ref{xy}) and (\ref{xygh}), the steering criterion (\ref{th3}) has the same ability of detecting steering as (\ref{steerxygh}).

In order to compare the effectiveness of our steerability criteria with the existing ones, let us consider detailed examples.
We take the generators of $\left\{G_i\right\}$ (similarly for $\left\{H_i\right\}$ with $d_A$ replaced with $d_B$) to be
$\left\{G_i\right\}=\left\{E_k, E_{k, l}^{+}, E_{k, l}^{-}\right\}$, where
$$
\begin{aligned}
E_k  =\sqrt{\frac{1}{(k+1)(k+2)}}(\sum_{j=0}^k& |j\rangle\langle j|-(k+1)|k+1\rangle\langle k+1|),\\ &\ k=0,1, \ldots, d_A-2, \\
E_{k, l}^{+} =\frac{1}{\sqrt{2}}(|k\rangle\langle l|+| l\rangle\langle k|), &~ \ k<l, k, ~l=0,1, \ldots, d_A-1, \\
E_{k, l}^{-}  =\frac{-i}{\sqrt{2}}(|k\rangle\langle l|-| l\rangle\langle k|), &~\ k<l, k, ~l=0,1, \ldots, d_A-1.
\end{aligned}
$$

{\textit{Example 1.}}
Consider symmetric two qubit state,
\begin{equation}
\rho=p|\psi^{-}\ra\la\psi^{-}|+(1-p)|0\ra\la0|\otimes \frac{\mathbb{I}_2}{2},
\end{equation}
where $|\psi^{-}\ra=(|01\ra-|10\ra)/\sqrt{2}$, $\mathbb{I}_2$ denotes the $2\times 2$ identity matrix. In Ref.~\cite{PhysRevLett.116.090403}, Moroder $et\ al.$ shew that the state $\rho$ is steerable in both directions for $p>0.57735027$. Lai and Luo in Ref.~\cite{lai2023steerability} obtained that $\rho$ is steerable from Alice to Bob for $p> 0.55895936$
 and from Bob to Alice for $p > 0.50007805$. Following Theorem~\ref{th1}, we get that the state is steerable from Alice to Bob for $p>0.55825996$ by taking $\alpha=64.8$ and $\beta=38.7$, and steerable from Bob to Alice for $p >0.50000341$ by taking $\alpha=67.7$ and $\beta=135.4$. Following Theorem~\ref{th2}, we get that the state is steerable from Alice to Bob for $p>0.55825769$ by taking $\alpha=64.8$, $\beta=38.7$, $\eta=0.02$ and $\gamma=0.02$, and steerable from Bob to Alice for $p> 0.50000004$ by taking $\alpha=67.7$, $\beta=135.4$, $\eta=0.1$ and $\gamma=0.1$. Thus, our results of Theorem~\ref{th1} and Theorem~\ref{th2} detect more steerable states in both directions and outperform the results of Refs.~\cite{PhysRevLett.116.090403} and~\cite{lai2023steerability} in this case, see Table \ref{table2}.

 \vspace{-1em}
\begin{table}[htp]
\caption{Detection of steerability for different criteria}
\begin{center}
\begin{tabular}{ccc}
\hline
Criteria&From Alice to Bob&From Bob to Alice\\
\hline
Ref.~\cite{PhysRevLett.116.090403}&$p>0.57735027$&$p>0.57735027$\\
Ref.~\cite{lai2023steerability}&$p> 0.55895936$&$p > 0.50007805$\\
Our Theorem 1&$p>0.55825996$&$p>0.50000341$\\
Our Theorem 2&$p>0.55825769$&$p>0.50000004$\\
\hline
\end{tabular}
\end{center}
\label{default}
\end{table}%
 \vspace{-2em}
\begin{table}[htp]
\caption{Detection of steerability for different criteria}
\begin{center}
\begin{tabular}{cc}
\hline
Criteria&Steerable in both sides\\
\hline
Ref.~\cite{PhysRevLett.116.090403}&$p>0.65612908$\\
Ref.~\cite{lai2023steerability}&$p>0.61882600$\\
Our Theorem 1&$p>0.61882600$\\
Our Theorem 2&$p>0.61882576$\\
\hline
\end{tabular}
\end{center}
\label{table2}
\end{table}%

{\textit{Example 2.}} Let us consider asymmetric two qubit state
\begin{equation}
\rho=p|\psi\ra\la\psi|+(1-p)\frac{\mathbb{I}_4}{4},\qquad 0\leq p\leq 1,
\end{equation}
where $|\psi\ra=\frac{2}{3}(|00\ra+|11\ra)+\frac{1}{3}|10\ra$, $\mathbb{I}_4$ denotes the $4\times 4$ identity matrix. From the positive partial transpose criterion this state is entangled if and only if $p>\frac{9}{25}$ \cite{PhysRevLett.77.1413}. In Ref.~\cite{PhysRevLett.116.090403}, Moroder $et\ al.$ shew that the state $\rho$ is steerable in both directions for $p>0.65612908$. Lai and Luo in Ref.~\cite{lai2023steerability} obtained that $\rho$ is steerable in both directions for $p > 0.61882600$. Our Theorem~\ref{th1} can obtain the same steerable interval when choosing $(\alpha,\beta)=(98,55)$ for Alice steering Bob and $(\alpha,\beta)=(55,98)$ for Bob steering Alice. Our Theorem~\ref{th2} shows that the state is steerable state in both directions for $p >0.61882576$ by taking $(\alpha,\beta,\eta,\gamma)=(98,55,0.1,0.1)$ for Alice steering Bob and $(\alpha,\beta,\eta,\gamma)=(55,98,0.1,0.1)$ for Bob steering Alice. We can conclude that Theorem~\ref{th1} and \ref{th2} detect more steerable states than the result of Ref.~\cite{lai2023steerability} in this case, see Table \ref{table2}.

\section{Conclusion}

By employing parameterized correlation matrices, we have obtained families of steerability criteria via generators of the special unitary group $SU(d)$, with respect to the steerability criteria based on the Heisenberg-Weyl observable basis given by Lai and Luo in Ref.~\cite{lai2023steerability}. By detailed examples we have shown that our criteria may detect more steerable states by choosing appropriate parameters. Our approach for arbitrary dimensional bipartite systems may be also applied to multipartite systems.

\bigskip
\noindent{\bf Acknowledgments}\, \,
This work is supported by the National Natural Science Foundation of China (NSFC) under Grants 12075159 and 12171044, Beijing Natural Science Foundation (Grant No. Z190005), Academician Innovation Platform of Hainan Province, and Changsha University of Science and Technology (Grant No. 097000303923).

\bigskip

\noindent{\bf CRediT authorship contribution statement}\, \,
Qing-Hua Zhang: Conceived the study, Wrote the manuscript, Reviewed and critically revised the manuscript.
Lemin Lai: Wrote the manuscript, Reviewed and critically revised the manuscript.
Shao-Ming Fei: Wrote the manuscript, Reviewed and critically revised the manuscript.

\bigskip

\noindent{\bf Declaration of competing interest}\, \,
The authors declare that they have no known competing financial interests or personal relationships that could have appeared to influence the work reported in this paper.

\bigskip

\noindent{\bf Data availability}\, \,
All data generated or analyzed during this study are included in the article.

\bibliographystyle{apsrev4-2}
\bibliography{zqhsteering}

\end{document}